\documentclass{article}%
\usepackage{amsmath}
\usepackage{chicago}
\usepackage{amsfonts}
\usepackage{amssymb}
\usepackage{graphicx}%
\begin{document}

\title{On Wave Function Representation of Particles as Shock Wave Discontinuities}
\author{Babur M. Mirza\\Department of Mathematics, \\Quaid-i-Azam University, 45320 Islamabad. Pakistan.\\E-mail: bmmirza2002@yahoo.com}
\maketitle

\begin{abstract}
In quantum theory particles are represented as wave packets. Shock wave
analysis of quantum equations of motion shows that wave function
representation in general and wave packet description in particular contains
discontinuities due to a non-zero quantum force. The quantum force causes wave
packet dispersion which results in the intersection of characteristic curves
developing a shock discontinuity. Since quantum force vanishes for localized
quantum density waves [1], it is thus established that localized quantum
density waves form the only class of wave function representation of particles
in quantum theory without shock wave discontinuities.

\end{abstract}

\section{Introduction}

Wave-particle duality is a central aspect of quantum processes. It is well
known that the Schr\"{o}dinger equation describes such a duality in the form
of wave packets solutions, particularly with the Gaussian wave packets [2-5].
Wave packet solutions of the Schr\"{o}dinger equation possess an interesting
properly that even in the absence of an external potential they exhibit
dispersion. Such dispersion occurs at very short time scales, causing
interference for electron waves.

In quantum potential formalism, wave packet dispersion results from the
tendency of quantum trajectories to accelerate away from each other. Quantum
trajectories for a free Gaussian wave packet, given by $x(t)=u_{0}t+x_{0}%
\sqrt{1+(\hbar t/2m\sigma_{0}^{2})^{2}}$, indicate that components of the wave
packet starting off at close by yet different initial positions and initial
speeds intersect after a short time, exhibiting a shock formation.
Intersection of characteristics lines is a typical aspect of nonlinear
systems. However, in quantum dynamics of wave packets, quantum trajectories
are assumed to be non-crossing. In the following exact analysis of quantum
wave packet dynamics we drop this assumption, and the complete nonlinear
system is analyzed using general theory of Riemann invariants. This leads to
the interesting result that quantum wave packets develop shock wave
discontinuities immediately after their formation.

Beginning with the next Section, we investigate the existence of shock wave
phenomenon in quantum hydrodynamic formulation of the Schr\"{o}dinger wave
equation. It is shown that for a general wave function solution of the
Schr\"{o}dinger equation a non-zero quantum force causes characteristics to
intersect, hence generates shock wave discontinuities in a quantized system.
Such discontinuities have a travelling wave form, and correspond to particle
motion in the free particle case. We take the example of a Gaussian wave
packet to calculate the position and time of formation of quantum shocks in
Section 3, whereas Section 4 gives a summary of the main conclusions of the
work and its relation to some recent experiments on electron waves.

\section{Shock Wave Analysis}

In the quantum potential approach, the general form of the wave function can
be written as $\psi(\mathbf{r},t)=R(\mathbf{r},t)\exp iS(\mathbf{r},t)/\hbar$.
Then the time-dependent Schr\"{o}dinger equation, with an external potential:
$i\hbar\partial\psi(\mathbf{r},t)/\partial t=-(\hbar^{2}/2m)\nabla^{2}%
\psi(\mathbf{r},t)+V(\mathbf{r})\psi(\mathbf{r},t) $ gives, after separating
real and imaginary parts, the following equations%
\begin{equation}
\frac{\partial\rho(\mathbf{r},t)}{\partial t}+\mathbf{\nabla}\cdot
(\rho(\mathbf{r},t)\mathbf{v}(\mathbf{r},t))=0, \label{1}%
\end{equation}%
\begin{equation}
\frac{\partial\mathbf{v}(\mathbf{r},t)}{\partial t}+\left(  \mathbf{v}%
(\mathbf{r},t)\cdot\mathbf{\nabla}\right)  \mathbf{v}(\mathbf{r},t)=-\frac
{1}{m}\mathbf{\nabla}(V(\mathbf{r})+mQ(\mathbf{r},t)). \label{2}%
\end{equation}
where $Q(\mathbf{r},t)=-(\hbar^{2}/2m^{2}R(\mathbf{r},t))\nabla^{2}%
R(\mathbf{r},t)$, and $\rho(\mathbf{r},t)=$ $R(\mathbf{r},t)^{2}$. Equations
(1) and (2) are the basic equations of quantum dynamics [6] in a fixed
Eulerian frame, with respect to which the relative velocity of an element is
$\mathbf{v}(\mathbf{r},t)$.

We write the equation (1) and (2) as a single matrix equation, and keep to the
one dimensional case only. Then equations (1) and (2) give,%
\begin{equation}
\left[
\begin{array}
[c]{cc}%
1 & 0\\
0 & 1
\end{array}
\right]  \left[
\begin{array}
[c]{c}%
\rho_{t}\\
u_{t}%
\end{array}
\right]  +\left[
\begin{array}
[c]{cc}%
u & \rho\\
Q_{\rho} & u
\end{array}
\right]  \left[
\begin{array}
[c]{c}%
\rho_{x}\\
u_{x}%
\end{array}
\right]  =0. \label{3}%
\end{equation}
where $u$ is the component of the velocity $\mathbf{v}$ along the
$x$-direction. Here $\rho_{t}=\partial\rho/\partial t$ and $Q_{\rho}=\partial
Q/\partial\rho$, etc.

The eigenvalues and eigenvectors for system (3) can be calculated from\ the
characteristic equation%
\begin{equation}
\det\left[
\begin{array}
[c]{cc}%
u-\lambda & \rho\\
Q_{\rho} & u-\lambda
\end{array}
\right]  =0, \label{4}%
\end{equation}
which gives%
\begin{equation}
\left(  u-\lambda\right)  ^{2}-\rho Q_{\rho}=0. \label{5}%
\end{equation}
Thus the two eigenvalues $\lambda_{1,2}$ are%
\begin{equation}
\lambda_{1,2}=u\pm\sqrt{\rho Q_{\rho}}. \label{6}%
\end{equation}
According to Riemann theory of shock waves [7], these eigenvalues give the
characteristic speed for families of characteristics $C_{+}$ and $C_{-}$. Thus
we have%
\begin{align}
C_{+}  &  :\lambda=\frac{dx}{dt}=u+\sqrt{\rho Q_{\rho}}=\lambda_{1}%
,\label{7}\\
C_{-}  &  :\lambda=\frac{dx}{dt}=u-\sqrt{\rho Q_{\rho}}=\lambda_{2}, \label{8}%
\end{align}
and the corresponding eigenvectors are given by%
\begin{equation}
\left[
\begin{array}
[c]{c}%
M_{r}^{1}\\
M_{r}^{2}%
\end{array}
\right]  =\left[
\begin{array}
[c]{c}%
\pm\sqrt{\rho/Q_{\rho}}\\
1
\end{array}
\right]  . \label{9}%
\end{equation}

In general, equations for characteristic lines for the system can be written
as%
\begin{equation}
X_{\pm}(t,t_{0})=x(t_{0})+\left(  \frac{dx}{dt}\right)  _{t=t_{0}}(t+t_{0}),
\label{10}%
\end{equation}
where $\left(  dx/dt\right)  _{t=t_{0}}=\lambda_{r}(t=t_{0})$. For shocks to
develop, these characteristic (10) must intersect at some common point in
space. This occurs if the slope of each characteristic increases with the
initial time $t_{0}$. In view of equations (7) and (8) this is the case
provided $Q_{\rho}\neq0$, that is, if the quantum force $\partial Q/\partial
x$ does not vanish. Another way to state shock condition is to expand function
$X_{\pm}(t,t_{0})$ for $\delta t_{0}<<1$ in the neighborhood of each
characteristics as%
\begin{equation}
X_{\pm}(t,t_{0}+\delta t_{0})\approx X_{\pm}(t,t_{0})+\frac{\partial X_{\pm
}(t,t_{0})}{\partial t_{0}}\delta t_{0}, \label{11}%
\end{equation}
then any two neighboring characteristics $X_{\pm}(t,t_{0})$ and $X_{\pm
}(t,t_{0}+\delta t_{0})$ intersect provided%
\begin{equation}
\frac{\partial X_{\pm}(t,t_{0})}{\partial t_{0}}=0. \label{12}%
\end{equation}
This is the shock condition for the system (3), which we shall use in the
following to calculate the time of shock formation.

Having obtained the necessary (and sufficient) shock conditions (12), we can
now explicitly determine the (travelling) shock wave solutions for the system
(3). This is done in the Appendix, using Riemann invariants. These
calculations show that the shock waves develop if the quantum force is
non-zero, thus the phenomenon is of purely quantum nature. Since for a free
Gaussian wave packet the quantum force is non-zero in general, this implies
the existence of quantum shocks in the case of free Gaussian wave packets.

\section{Quantum Shocks for the Case of a Gaussian Wave Packet}

Gaussian wave packet solution to the Schr\"{o}dinger equation for a free
particle is given by the wave function%
\begin{equation}
\psi(x,t)=\frac{1}{\left(  2\pi s\right)  ^{3/4}}\exp\left(  ik(x-u_{0}%
t/2)-(x-u_{0}t)^{2}\right)  /4s\sigma_{0}, \label{13}%
\end{equation}
where $u_{0}$ is the uniform constant speed of the wave packet, and the
measure of the spread $\mid s\mid=$ $\sigma=\sqrt{\sigma_{0}^{2}(1+(\hbar
t/2m\sigma_{0}^{2})^{2})}$. Correspondingly, the amplitude and the phase
functions are given by
\begin{align}
R(x,t)  &  =\frac{1}{\left(  2\pi\sigma^{2}\right)  ^{3/4}}\exp-(x-u_{0}%
t)^{2}/4\sigma^{2},\label{14}\\
S(x,t)  &  =\frac{-3\hbar}{2}\tan^{-1}(\frac{\hbar}{2m\sigma_{0}^{2}%
}t)+mu(x-u_{0}t/2)+(x-u_{0}t)^{2}\frac{\hbar^{2}}{8m\sigma_{0}^{2}\sigma^{2}%
}t, \label{15}%
\end{align}
and the quantum potential is,%
\begin{equation}
Q(x,t)=\frac{\hbar^{2}}{4m^{2}\sigma^{2}}(3-\frac{(x-u_{0}t)^{2}}{2\sigma^{2}%
}). \label{16}%
\end{equation}
Using $u=\left(  \partial S/\partial x\right)  /m$, we obtain from equation
(16) the speed of a wave packet element%
\begin{equation}
u=u_{0}+\frac{\hbar^{2}t}{4m^{2}\sigma_{0}^{2}\sigma^{2}}(x-u_{0}t).
\label{17}%
\end{equation}
hence by integration, the position of a wave packet component is%
\begin{equation}
x(t)=u_{0}t+x_{0}\sqrt{1+(\frac{\hbar t}{2m\sigma_{0}^{2}})^{2}}. \label{18}%
\end{equation}
Here $x_{0}$ and $u_{0}$ denote the initial position and velocity of the wave
components, respectively.

Since by the above analysis the characteristics along which each wave packet
component travels must intersect, we determine the location of the quantum
shock in this case. Using equations (7) and (8), the equation of
characteristic is given by%
\begin{equation}
X_{\pm}(t,t_{0})=x(t_{0})+\left(  u\pm\sqrt{\rho Q_{\rho}}\right)  _{t=t_{0}%
}(t+t_{0}). \label{19}%
\end{equation}
Then from the shock condition (12), we have%
\begin{equation}
x^{\prime}(t_{0})+\frac{d\left(  u\pm\sqrt{\rho Q_{\rho}}\right)  _{t=t_{0}}%
}{dt_{0}}(t+t_{0})-\left(  u\pm\sqrt{\rho Q_{\rho}}\right)  _{t=t_{0}}=0.
\label{20}%
\end{equation}
Substituting for $Q$, $u$, and $x$ from equations (16), (17) and (18)
respectively, and then taking $t_{0}=0$ we have, after some simplification,%
\begin{equation}
t_{s}=\frac{-8m^{2}\sigma_{0}^{4}}{\hbar^{2}x_{0}}\left(  u_{0}\pm\sqrt
{\frac{\hbar^{2}}{16\sigma_{0}^{3}}}\right)  , \label{21}%
\end{equation}
and therefore, from equations (7) and (8),%
\begin{equation}
x_{s}=(u_{0}\pm\sqrt{\frac{\hbar^{2}}{4\sigma_{0}^{3}}})t_{s}. \label{22}%
\end{equation}
Equations (21) and (22) give the time and position for the quantum shock
associated with the Gaussian wave packet (13).

\section{Conclusions}

The analysis presented in this paper shows that if in any region of space the
quantum force tends to increase, a shock-like situation will develop. This
must be so, since slope of the characteristics (19) then increases, causing
characteristic curves to intersect. This was explicitly shown for the case of
Gaussian wave packets, where the quantum force is equal to $\hbar^{2}%
(x-u_{0}t)/4m\sigma(t)^{4}$; which although decreasing first, then increases,
and eventually attains a constant limit as $x\rightarrow\infty$ and $t$
$\rightarrow\infty$ (Fig. 1). This indicates that the quantum force causes the
wave packet to burst rather than spread smoothly.

If the relative velocity of the Gaussian wave packet and the lab frame
coincides, quantum shock occurs at time $t_{s}=\sqrt{4m^{2}\sigma_{0}%
^{13}/\hbar^{2}x_{0}^{2}}$, and position $x_{s}=\sqrt{m^{2}\sigma_{0}%
^{10}/x_{0}^{2}}$, travelling with the speed $x_{s}/t_{s}$. Equation (22)
shows that this speed differs from the classical formula by a constant
$\hbar/\left(  2\sigma_{0}^{3/2}\right)  $ due to wave packet dispersion.

A similar analysis can be applied to the case of Airy beams [8], and to the
recently observed leviton structures [9], \ where limits on electron
interferometry has been reduced to the attosecond scale [10,11]. For quantum
density soliton waves, representing particle-like localization, such
discontinuities do not form, since quantum force in this case is identically
zero. This result also has implications for the problem of equivalence
principle in quantum theory [12].

\begin{center}
\textbf{APPENDIX: Riemann Invariants and the Shock Wave Solution}
\end{center}

To obtain the Riemann invariants for the system (3), we take a linear
combination of the equations (3) with coefficients being the components of the
eigenvectors (9), this gives:%
\begin{equation}
\left(  \rho_{t}+u\rho_{x}+\rho u_{x}\right)  \pm\sqrt{\rho/Q_{\rho}}\left(
u_{t}+Q_{\rho}\rho_{x}+uu_{x}\right)  =0. \tag{A1}%
\end{equation}
Since $\left(  \partial\rho/\partial\eta\right)  /\left(  \pm\sqrt
{\rho/Q_{\rho}}\right)  =\left(  \partial u/\partial\eta\right)  $, it follows
that%
\begin{equation}
\frac{\partial u}{\partial\rho}=\pm\sqrt{\frac{Q_{\rho}}{\rho}}\text{, and
}\frac{\partial\rho}{\partial u}=\pm\sqrt{\frac{\rho}{Q_{\rho}}}. \tag{A2}%
\end{equation}
Then substituting for $u_{x}=u_{\rho}\rho_{x}=\pm\sqrt{Q_{\rho}/\rho}\rho
_{x},$ and $\rho_{x}=\rho_{u}u_{x}=\pm\sqrt{\rho/Q_{\rho}}u_{x},$ into
equation (A1) we have%
\begin{equation}
\left(  \rho_{t}+\left(  u\pm\sqrt{\rho Q_{\rho}}\right)  \rho_{x}\right)
\pm\sqrt{\rho/Q_{\rho}}\left(  u_{t}+\left(  u\pm\sqrt{\rho Q_{\rho}}\right)
u_{x}\right)  =0, \tag{A3}%
\end{equation}
and using equations (7) and (8):%
\begin{equation}
d\left(  u\pm F(\rho\right)  )=0,\text{ where }F(\rho)=\int\sqrt{Q_{\rho}%
/\rho}d\rho\text{.} \tag{A4}%
\end{equation}
Integrating we have the two constants (Riemann invariants), $A$ and $B$ along
the characteristics:%
\begin{align}
u+F(\rho)  &  =A(\xi),\text{ along }\Gamma_{1}\text{ with parameter }%
\xi=x+\lambda t,\tag{A5}\\
u-F(\rho)  &  =B(\eta),\text{ along }\Gamma_{2}\text{ with parameter }%
\eta=x-\lambda t. \tag{A6}%
\end{align}
Thus eliminating $u$ and $F(\rho)$, we have the solution for $u(x,t)$ and
$\rho(x,t)$:
\begin{align}
u(x,t)  &  =A(x+\lambda t)+B(x-\lambda t),\tag{A7}\\
F(\rho)  &  =A(x+\lambda t)-B(x-\lambda t). \tag{A8}%
\end{align}
which can be easily verified as a (traveling) shock wave solution to the
system (3). We notice in the above analysis that if the quantum force is zero,
then the shock speeds $\lambda_{1}$ and $\lambda_{2}$ equal the speed $u$.

\bigskip

Figure Caption:

Fig. 1: Quantum force for the Gaussian wave packet (13), with $\hbar$, $m$,
$\sigma_{0}$ unity, and $u_{0}=10$ units.

\begin{thebibliography}{99}                                                                                               %


\bibitem {[1]}B. M. Mirza, Mod. \textit{Phys. Lett.} B \textbf{28} 1450253
(2014) .

\bibitem {[2]}N. Lee, et.al., \textit{Science }\textbf{332}, 330 (2011).

\bibitem {[3]}V. Krueckl and T. Kramer, 2009 \textit{New J. Phys}.
\textbf{11}, 093010 (2009).

\bibitem {[4]}J. Billy, et al., \textit{Nature} \textbf{453}, 891 (2008).

\bibitem {[5]}R. E. Wyatt , C. J. Trahan, \textit{Quantum Dynamics with
Trajectories} (Springer-Verlag, New York, 2005).

\bibitem {[6]}V. E. Madelung, \textit{Z. Physik}. \textbf{40}, 322 (1926); L.
de Broglie, \textit{Electrons et Photons, Report au Ve Conseil Physique
Solvay} (Gauthier-Villiars, Paris, 1930); D. Bohm, \textit{Phys. Rev}.
\textbf{85}, 166 (1952). \ \ \ \ \ \ 

\bibitem {[7]}R. Cournat and R. Friedrichs, \textit{Supersonic Flows and Shock
Waves} (Springer-Verlag, New York, 1985). \ 

\bibitem {[8]}N. Voloch-Bloch, et.al., \textit{Nature} \textbf{494}, 331 (2013).

\bibitem {[9]}J. Dubois, et al.\textit{\ Nature}\ \textbf{502}, 659 (2013). \ 

\bibitem {[10]}M. Holler, et.al., \textit{Phys. Rev. Lett.} \textbf{106},
123601 (2011).

\bibitem {[11]}T. Remetter, et.al., \textit{Nature Physics} \textbf{2}, 323 (2006).

\bibitem {[12]}B. M. Mirza, to appear.
\end{thebibliography}
\end{document}